# Promoting Diversity and Inclusion in Astronomy Graduate Education: an Astro2020 Decadal Survey White Paper

**AAS Task Force on Diversity and Inclusion in Astronomy Graduate Education**

**Endorsed by the AAS Board of Trustees**

Endorsing Task Force Members
Alexander Rudolph, Cal Poly Pomona (co-Chair)
Gibor Basri, UC Berkeley (co-Chair)
Marcel Agüeros, Columbia Univ.
Ed Bertschinger, MIT
Kim Coble, San Francisco State Univ.
Megan Donahue, Michigan State Univ.
Jackie Monkiewicz, Arizona State Univ.
Angela Speck, Univ. of Missouri
Rachel Ivie, AIP
Christine Pfund, Univ. of Wisconsin-Madison
Julie Posselt, Univ. of Southern California

Additional Endorsers (Task Force Working Group Members)
Peter Frinchaboy, Texas Christian Univ.
Jenny Greene, Princeton Univ.
Emily Levesque, Univ. of Washington
Seth Redfield, Wesleyan Univ.
Mariangelly Diaz-Rodriguez, Florida State Univ.
David Helfand, Columbia Univ.
Eric Hooper, Univ. of Wisconsin
Richard Anantua, UC Berkeley
Jarita Holbrook, University of the Western Cape, South Africa
Douglas Richstone, Univ. of Michigan
Meg Urry, Yale Univ.

# 1. Introduction

## 1.1 Purpose of this White Paper

The purpose of this white paper is to provide guidance to funding agencies, leaders in the discipline, and its constituent departments about strategies for (1) improving access to advanced education for people from populations that have long been underrepresented and (2) improving the climates of departments where students enroll. The twin goals of improving access to increase diversity and improving climate to enhance inclusiveness are mutually reinforcing, and they are both predicated on a fundamental problem of inequality in participation.

This white paper is structured as follows: below we provide some context for our original report, which was submitted to the American Astronomical Society (AAS) and endorsed by its Board of Trustees at its January 2019 meeting. We then reproduce the recommendations from that report in Section 2, which are directed to individual departments and to the AAS. We do not include the supporting materials behind these recommendations as they can be found in the full report, which can be found here (Rudolph et al. 2019). This white paper includes specific, direct recommendations to the funding agencies in Section 3. We also recommend that funding agencies support the recommendations in the main Task Force report.

## 1.2 National Context

According to the latest statistics from the National Science Foundation (NSF 2015), underrepresented minority (URM)[1] students made up only 3% of PhDs in astronomy between 2002-2012, yet they comprise 30% of the general population. There was a total of 4±2 URM PhDs per year in astronomy during that period nationally – a percentage and a number that are unacceptable and, if astronomy is to succeed as a largely publicly funded and publicly supported endeavor in America, ultimately unsustainable. The 2010 Decadal Survey of Astronomy highlighted this problem, noting that "Little progress has been made in increasing the number of minorities in Astronomy," and recommending "Partnerships of community colleges and minority-serving institutions [MSIs] with research universities and with national centers and laboratories" to overcome this underrepresentation. Little has changed since that report was published.

The inequalities we are looking to address are not new problems for the field, nor are we the first to engage with them. Reports from the National Academies of Sciences, Engineering, and Medicine (NASEM) in 2007, 2011, and 2018 each called attention to the fact that the available talent pool in US STEM is not being utilized and this has been the case for at least 50 years – with rather little progress (NASEM 2007; 2011; 2018a; 2018b). Efforts in astronomy to date, while making a difference in specific places and making a small dent in specific problems, are not moving the needle quickly enough at the field level, particularly because the nation's demographics are evolving to make patterns of underrepresentation (and therefore talent underutilization) even worse. For example, the fastest growing racial/ethnic group in the US is Hispanic/Latinx, and although people identifying as such have seen impressive gains in overall academic preparation and college attendance, their rates of PhD attainment and representation in the professoriate lag well behind their share of the current population.

Graduate education is a crucial place for intervention, for without graduate degrees, astronomers cannot be a part of the research enterprise. It is a crucial part of the opportunity structure in the discipline. The 2011 NASEM report, *Expanding Underrepresented Minority Participation: America's Science and Technology at the Crossroads* named the "transition to graduate study" as one of two key action areas (NASEM 2011). The most recent report, *Graduate STEM Education for the 21st Century* acknowledged the need for a systems approach to improving graduate education, and the commitment of a broad group of stakeholders in the scientific enterprise; its recommendations include a suite of actions to make STEM graduate education learning environments more equitable, diverse, and inclusive (NASEM 2018a). Our recommendations thus align well with those that NASEM has made.

The fact that the population of first-year graduate students so poorly reflects the population of bachelor's degree recipients implies that barriers exist in recruitment and admissions processes. And if we consider both the academic and demographic characteristics of those who start PhD programs, and assess who ultimately completes the degree, the need to improve PhD programs' mentoring and retention efforts

---

[1] Underrepresented minorities (URM) categories include Black/African American, Latinx/Hispanic, Native American/American Indian, Alaskan Native, and Pacific Islander

becomes clear. In short, the best available research and data compel a need to look beyond student characteristics to the operation and climate of institutions. Improving graduate education means we need to take a hard look at institutional priorities and their processes of assessing who is qualified, can contribute, and ultimately belongs. In the end, encouraging diversity and inclusion means thinking holistically about student potential and who belongs in the discipline. In many places, a conversation about the imperative to address racial inequalities has also called attention to the need to make our communities more inclusive of other marginalized groups. This task force has tried to address diversity with respect to race/ethnicity, gender identity, sexual orientation, and neurodiversity and disability status. Rather than focusing on a particular type of "perfect" or "ideal" student or particular processes of admission or retention, we should promote practices that will enable all of us to see and develop talent and potential more broadly than we have before, so as not to miss potential contributors to the field of astronomy.

**1.3 Prior Studies of Graduate Education in Astronomy**

Twenty years ago, a report on graduate education, entitled *The American Astronomical Society's Examination of Graduate Education in Astronomy* was jointly created by the AAS Education Policy Board and Graduate Advisory Board (AAS 1996). The primary goal of that report was to address the perceived overproduction of PhDs at a time of great funding uncertainty. Its three key recommendations were:

1. Define and Support Experiments to Enrich Graduate Education
2. Re-examine the Master's Degree in Astronomy
3. Provide Students with the Information and Experience Necessary to Make Informed Career Decisions

A great deal of effort went into producing that report, and much of what it found and recommended is still relevant today. As such, any effort to assess graduate education today should be informed by the findings of that report. The one explicit mention of diversity in the report occurred in Section 4.1.3 entitled, *Deliberate reduction of the population of graduate students or of graduate departments is not wise*:

> "…it was agreed that the admissions process is imperfect; identifying college seniors with the combination of intelligence and temperament matched to a research career is, with few exceptions, extremely difficult. Practicing "birth control" at this stage would result in premature evaluations based more on "objective" criteria than on assessment of a student's performance in a graduate research department. Moreover, "birth control" at this early stage would almost certainly compromise the ability of graduate departments to meet their stated goal of enhancing diversity in the physical sciences."

This statement highlights a key point: graduate admissions criteria are one of the steepest barriers to increasing diversity in astronomy. Education and other social science research have shown that common uses of traditional measures of ability used in graduate admissions, particularly the general Graduate Record Exam (GRE) and Physics subject GRE (PGRE), disproportionately exclude groups who are already underrepresented (Miller and Stassun 2014; Posselt 2016; Steele and Aronson 1995). Further, these same measures are poor predictors of PhD completion and long-term success in research, the two main goals of most PhD programs (Miller et al. 2019; Petersen et al. 2018; Glanz 1996; Sternberg and Williams 1997; Helms 2009).

Admissions was one topic of the 2015 Inclusive Astronomy meeting, which produced an extensive report known as the Nashville Recommendations (2015). Our report continued efforts in that vein, and includes a number of recommendations for departments from the Nashville report that will lead to the improvement of recruitment, admissions, climate, mentoring, and retention. Increasing the uptake of the Nashville Recommendations was part of the motivation for this Task Force.

The AAS Council was the first disciplinary society to [recommend that its constituent PhD programs eliminate or make optional the GRE exam in graduate admissions](), and a number of major astronomy departments have recently voted to eliminate or make optional the PGRE in their graduate admissions requirements ([Astrobites 2016]). No single reform alone, however, will solve the diversity and inclusion problem. Graduate faculties need to examine their entire programs because to consider how, through Admissions and financial aid decisions, curriculum requirements and qualifying processes, as well as mentoring and support structures (or lack thereof) work together to enable or suppress diversity and inclusion in graduate programs and ultimately in the field overall. Therefore, graduate faculties need to examine their entire programs to assess their contributions to the lack of diversity and inclusion in astronomy. Once these are identified, graduate programs must act to remedy these problems. In addition, funding agencies need to evaluate their decision-making process to assure that it aligns with and supports departments in achieving these goals of diversity and inclusion in the field.

In most departments, there are both leaders for diversity and those who have resisted changes to this end. However, as a field, astronomy has also been a leader among STEM disciplines in encouraging both grass-roots and top-down efforts to improve admissions. Many of the top-tier astronomy PhD-producing institutions – almost none of which are MSIs – have begun sincere and genuine efforts to improve access to their graduate programs for URMs, women, and other underrepresented groups (such as LGBTIQA* and the differently-abled). Other notable efforts have included the creation of "bridge programs" and novel summer research programs designed to reach large numbers of URM students who do not traditionally participate in Research Experience for Undergraduate (REU) programs. Furthermore, professional societies aligned with AAS have also moved to address the issue of diversity and inclusion in graduate education: notably, the American Physical Society (APS) has convened a [national meeting](#) on physics graduate education and bridge programs and has piloted a network for access and inclusion in graduate education through an [NSF-INCLUDES award](#). It is time for the funding agencies, the AAS, and astronomy departments nationally to formally engage in these efforts.

Meanwhile, the three principal recommendations of the 1996 report remain relevant. The connections between our discipline and the wider economy have multiplied in the past twenty years: from robotics to statistical inference, and from design to big data, the prospects for an astronomy PhD student outside the traditional academic track have brightened. Both our curricula and our cultural attitudes need adjustments to reflect this new reality. Furthermore, many potential future members of the astronomical community – perhaps especially URM students, who disproportionately face difficult economic circumstances – need viable options for employment and professional development other than a traditional doctorate leading to an academic position, as well as viable career "off-ramps" if they do not complete the PhD for any reason.

Finally, we see a continuing need to provide prospective students with complete and accurate information about graduate program opportunities, climates, and outcomes. The 1996 report contained explicit recommendations for offering such information to prospective students, recommendations on which action has been minimal. It is our hope that more explicit, coordinated actions will be taken to this end. Indeed, the AAS has coordinated the postdoc market in astronomy by imposing uniform decision deadlines and creating a universally used job register. A similar initiative for the graduate education market could likewise develop an enforceable set of community standards for the provision of information to prospective students.

**1.4 Theory of Change**

Underlying the recommendations in this white paper is current organizational and social theory about why and how large, distributed organizations change. Collectively, the strategy that we outline consists of recommendations to funding agencies and the AAS to take actions that will motivate astronomy PhD programs to adopt equitable, inclusive practices and climate as well as recommendations to individual departments. The funding agency recommendations are designed to use the central role funding has in the structure and function of graduate astronomy departments, and to encourage the funding agencies to fund activities that support diversity and inclusion practices as well as to ensure that the selection criteria fully support such practices. The AAS recommendations are for top-down changes that include measuring the climate and other characteristics of astronomy PhD programs, investing in the development and advancement of evidence-based practices, and recognizing departments that adopt such practices. Alongside the top-down change that funding agencies and the AAS have the leverage to encourage, we also make recommendations to astronomy departments, to encourage change from the bottom up.

The recommendations herein and the actions we hope will follow continue a decades-long process of improving the discipline's inclusiveness. AAS has been, at varying points in time, more or less engaged in that effort. The Women in Astronomy and Inclusive Astronomy meetings, in addition to the work of the 1996 AAS Education Policy Board and the Graduate Advisory Board, have each issued reports and recommendations for the field that hold implications for the policies and practices that shape access, equity, diversity, and inclusion in graduate programs. Funding agencies have also had varying levels of engagement: highlights include the Partnerships for Astronomy and Astrophysics Research and Education (PAARE) program, a highly successful investment in numerous programs (e.g., Fisk-Vanderbilt Master's-to-PhD Bridge program, Columbia Bridge program, Cal-Bridge/CAMPARE, AstroComNYC, etc.) that have shown significant success at increasing the numbers of URM and women students pursuing and obtaining PhDs in astronomy and other STEM fields.

What will it take to catalyze widespread adoption of these recommendations? We think that at least four major factors have been missing, and our report addresses them directly:

1) Although the various funding agencies have shown, at times, a willingness to support programs, such as PAARE, that promote diversity and inclusion in astronomy, they have not made diversity and inclusion central parts of their funding strategies in ways that would lead to sustained funding needed to incentivize true cultural changes in the astronomy community

2) The absence of a coordinated data collection effort with standard metrics has prevented departments from making meaningful comparisons – both with prior versions of themselves (to benchmark their progress) and with other astronomy departments (to gauge their equity and inclusiveness relative to that of the field and/or peer departments). Data of various types provide a mirror through which departments – and the field – can see themselves more clearly. Therefore, underlying both the top-down (i.e., agency-driven) and bottom-up (i.e., department-driven) efforts will be an ongoing conversation about data and evidence, and an entire section (2.3) of the main report is dedicated to the data that progress demands

3) Clarity and evidence have been missing about the practices that hold potential to move the needle on equity, diversity, and inclusion. Although there was widespread support in principle for the Nashville Recommendations, some departments struggled to adopt them in the absence of strong evidence at the time for their effectiveness. Therefore, an important part of this report – and the Task Force's composition – was to bring the best current evidence and research to bear in making the case for inclusive practices. It is important to note that although the research base for graduate education is growing quickly, it remains much smaller than that for K-12 and undergraduate education. Therefore, although we bring current research to bear on our recommendations wherever possible, we also offer examples from individual and small groups of departments throughout the field, recognizing that these individual cases and stories do not permit the same generalizability that research offers – and which faculty in a data-driven field like astronomy may yet need to be persuaded that a given recommendation will be effective

4) Although astronomy has been a leader in grassroots efforts of individual faculty and departments, widespread adoption of inclusive practices will also be encouraged by institutionalizing systems and to incentivize good behavior. Research from the LEED[2] certification system for environmental stewardship in building design and construction has demonstrated the potential of centralized recognition systems for motivating socially conscious organizational behavior. [Athena SWAN and the Race Equality Charter](#) take a similar approach in the United Kingdom to recognize universities adhering to practices that encourage gender and race equity, and a similar system, the STEM Equity Achievement (SEA) Change system, is under development here in the United States. It uses self-assessment, common metrics, and public certification to recognize American universities for transformations aligned with goals of equity, diversity, and inclusion

To summarize, we believe the best way for astronomy to make progress as a field toward diversity and inclusion is through a combination of top-down actions by funding agencies and AAS, and bottom up actions by departments. Diversity and inclusion should be not only our goals, but also principles to embody in the change process. We hope that the structure of our recommendations – and our attention to data, research evidence, and the need for incentives – catalyzes widespread adoption of practices that have already gained support across the field, by improving the data environment, by bringing clarity about what the most promising practices are, and by creating a system that motivates desired behaviors. Collectively, following these recommendations holds potential to both restructure the system by which students gain access to graduate education in astronomy and improve the climates in PhD programs and the field.

---

[2] Leadership in Energy and Environmental Design (LEED)

**2. Goals and Recommendations from the Task Force Report**

**A. Admissions: Goals and Recommendations to Departments**

**Goals**
   A. The demographics of students admitted to PhD programs in astronomy should reflect those of the availability pool at the undergraduate level
   B. Admissions criteria and processes should be designed to broaden the definitions of excellence and merit to create greater diversity in admitted cohorts
   C. Applying to a graduate program should be a transparent, informed process

**Recommendations to Departments**
   1. Partner with and recruit from undergraduate programs that produce large numbers of graduates from underrepresented groups (e.g., MSIs, HSIs[3], and Tribal Colleges)
   2. Implement evidence-based, systematic, holistic approaches to graduate admissions, based on the existing literature as well as on self-study when possible
   3. Coordinate with graduate schools and other campus offices to ensure that program level policies and practices aimed at diversity and inclusion are supported and amplified at the institutional level

**B. Retention: Goals and Recommendations to Departments**

**Goals**
   A. End harassment and bullying in and around astronomical workplaces
   B. Provide an accessible environment, including but not limited to full ADA[4]-compliance
   C. Provide a healthy, welcoming, family-friendly environment
   D. Provide effective mentoring through evidence-based practices and expanded networking opportunities
   E. Adopt teaching and learning practices that support all students, especially those with marginalized identities

**Recommendations to Departments**
   1. Engage in genuine, open, and sometimes difficult conversations
   2. Conduct assessments to identify areas of need or opportunities
   3. Create short- and long-term actionable department plans with measurable outcomes that address the five goals
   4. Incentivize and support professional development in the support of the five goals
   5. Take actions based on the departmental plan and monitor progress toward outcomes, employing inclusive processes
   6. Encourage ongoing improvements toward inclusiveness by iterating through the process represented in steps 1-5

**C. Data Collection and Metrics for Success: Goals and Recommendations to Departments**

**Goals**
   A. Measure progress toward the recommendations regarding Admissions and Retention

---

[3] Minority Serving Institutions (MSIs); Hispanic Serving Institutions (HSIs)

[4] Americans with Disabilities Act (ADA) of 1990, a civil rights law that prohibits discrimination based on disability

B. Measure trends in field-wide demographic and climate data to assess which practices are effective and for whom
C. Help departments advance their goals for diversity, equity, and inclusion using data and metrics for success

**Recommendations to Departments**
1. Participate in the recommended AAS/AIP[5] national demographic and climate survey, and encourage all relevant members (e.g., graduate students, postdocs, researchers, faculty) to participate
2. Regularly collect and analyze data relevant to graduate education, including the demographics of applicant pools, admitted and enrolled students, and disaggregated progress and success rates
3. Assess the success of steps taken to improve the educational experience of graduate students using an evidence-based rubric
4. Report results on progress in implementing the recommendations of this Task Force on the platform provided by the AAS and on departmental websites

**D. Goals and Recommendations to the AAS**

**Goals**
A. Measure the status and progress of diversity and inclusion in programs producing graduate degrees in astronomy
B. Provide a platform that incentivizes, recognizes, and disseminates steps that these programs take to increase diversity and inclusion in astronomy
C. Actively participate in the effort to produce, test, and disseminate new promising practices that increase diversity and inclusion in astronomy

**Recommendations to AAS**
1. Partner with the AIP Statistical Research Center to collect demographic and climate data
2. Recruit departments to adopt the recommendations of this Task Force
3. Create a platform for encouraging departments to adopt best practices and to track their adoption over time
4. Invest in the continued development, sharing, and curation of research- and best-practice-based toolkits that enable graduate programs to implement evidence-based recruitment, admissions, and mentoring practices
5. Encourage participation by the AAS equity committees and working groups in the AAAS[6] SEA Change[7] initiative

**3. Goals and Recommendations of the Task Force to Funding Agencies**

**Goals**
A. Provide on-going, long-term funding support to programs showing promise and results toward promoting diversity and inclusion in astronomy
B. Provide funding mechanisms for departments and other groups to implement best practices in promoting diversity and inclusion as outlined in the Task Force report
C. Provide funding support for research into best practices that promote diversity and inclusion in astronomy
D. Provide meaningful incentives to grant applicants to make diversity and inclusion a priority

---

[5] American Astronomical Society (AAS) and American Institute of Physics (AIP)

[6] American Association for the Advancement of Science (AAAS)

[7] Science Technology Engineering and Science (STEM) Equity Achievement (SEA) Change

**Recommendations to Funding Agencies (listed in priority order)**

1. Fund programs such as the highly successful PAARE, MUCERPI,[8] and FaST[9] programs, that focus on providing entry into the profession to underrepresented populations, especially at the post-secondary levels. Over the past two decades, these programs, at times funded at the level of less than ~1% of the respective directorate/office budgets, have been responsible for the greatest leaps in progress for diversifying astronomy and astrophysics, but have now all become defunded
    a. Grant funding for programs supporting diversity and inclusion should be at least a few percent of each directorate's budget (NSF AST, NASA Astrophysics, and DOE Office of Science). We note that providing the capacity for funding these targeted programs requires maintaining robust support for standard grants, as funding for diversity, equity, and inclusion (DEI) programs is frequently squeezed out of agency budgets when individual investigator grant budgets are threatened
    b. Utilize accomplishment-based renewal or other such evaluation mechanisms to ensure that deserving programs receive ongoing long-term support

2. Dedicate funding at the mission level (e.g., space-based great observatories and explorers, large ground-based facilities, etc.) to broadening participation, such as supporting graduate students, partnerships with minority serving institutions, postdoctoral fellowships, etc. This ensures development of a future diverse workforce specifically connected to the missions/facilities that will be the state of the art for the coming generation
    a. At least ~1% of the mission/facility development/operating budget should be committed to broadening participation
    b. Cooperative agreements to management organizations for facilities should explicitly provide funds for engaging diverse groups in the work, mission, and future workforce development of those facilities

3. Provide funding to departments to implement the AAS Task Force recommendations in promoting diversity and inclusion in astronomy/astrophysics graduate education
    a. Dedicated funding should be made available to astronomy/astrophysics PhD-granting departments who choose to participate in the AAS Task Force recommendations described above. Commitment to these recommendations may require a considerable investment in time, money, and personnel, which may become a limiting factor for resource-strapped institutions. Dedicated funding removes a potential barrier to participation for many departments
    b. Funding for departments at MSIs/HSIs/master's-serving institutions to engage in departmental transformation via communities of practice, professional development, curriculum implementation and other best practices to support equity and inclusion. These departments provide the diverse population of students who can pursue the pathways to a PhD opened up by PhD-granting departments implementing the AAS Task Force recommendations

4. Provide funding mechanisms to research best practices in promoting diversity and inclusion in graduate education and fund data gathering of the metrics as described in the AAS Task Force report. We estimate that funding is needed for these initiatives at the level of approximately $500,000 per year
    a. Fund at least one effort each year to measure success rates of DEI initiatives in astronomy specifically. The research described in understanding-interventions.org should be used as exemplars. Research about the success of

---

[8] Minority University College Education and Research Partnership Initiative in Space Science (MUCERPI); https://www.nasa.gov/home/hqnews/2003/sep/HQ_03302_minority_partners.html

[9] The Faculty and Student Teams (FaST) program supported 10-week summer research experiences jointly funded by the Department of Energy and the NSF. See https://haas.stanford.edu/students/cardinal-careers/fellowships/faculty-and-student-teams-fast-program for an example.

DEI initiatives in astronomy is needed because results may be different than in other fields
    b. Support the AAS in funding the data collection system described in the Task Force report. The AAS will need to set up a data collection infrastructure that allows collection of demographic and department climate data in order to measure progress toward the goal of full participation of under-represented groups in astronomy

5. Require graduate student mentoring plans in all NSF AST grants (in addition to the already required postdoc mentoring plans) and require graduate student and postdoc mentoring plans in all other agency grants that support either of these groups
    a. Require all science grants which fund graduate students to provide a graduate student mentoring plan, similar to the postdoc mentoring plan now required by the NSF. Such plans could consist of two parts: 1) a boilerplate section prepared by the applicant's home department describing their graduate student mentoring practices, and 2) a section for the applicant to outlined their personal mentoring plans. The department section would not be required and a poor department plan should not prevent a grant from being funded, but a strong department plan could enhance the applicant's personal mentoring plan. Proposals that showed a good integration between the two plans (department and individual) should be highly rated
    b. Weigh evaluation of these plans (both grad student and postdoc) significantly in the review process so that high-quality, inclusive mentoring becomes a high priority for grant applicants. Lack of an adequate graduate student and/or postdoc mentoring plan should potentially result in funding being denied, at the discretion of the program officer
    c. Contract with the social science community to develop rubrics for evaluating mentoring plans. Current panels lack the expertise to credibly evaluate these plans
    d. Provide professional development training to program officers in evaluating mentoring plans, and consider entrusting rating of such plans exclusively to the program officers, using rubrics developed under point 5c, leaving the scientific panels to evaluate what they can do best: science

6. Provide funding to professional societies to support professional development around equity and inclusion
    a. Support national and regional workshops, including continued support of the [New Faculty Workshop](#), which is currently sponsored by AAPT, AAS, APS, AIP, NSF, and Research Corporation for Science Advancement
    b. Fund professional development workshops at AAS meetings to support: graduate student mentoring, equity in teaching practices, and other strategies described in the Retention section of the Task Force Report